\documentclass[preprint,12pt]{elsarticle}

\usepackage{graphicx}
\usepackage{algorithm}
\usepackage[noend]{algpseudocode}
\usepackage{subcaption}

\usepackage{lineno}

\usepackage{amsmath,amsfonts,amssymb,amsthm}
\usepackage{mathtools}
\usepackage{commath}
\usepackage[sc,osf]{mathpazo}
\usepackage{url}

\journal{Computers \& Security}

\begin{document}

\begin{frontmatter}


\title{Random CapsNet Forest Model for Imbalanced Malware Type Classification Task}



\author{Aykut \c{C}ay{\i}r\corref{cor1}}
\ead{aykut.cayir@khas.edu.tr}

\author{U\u{g}ur \"{U}nal}
\author{Hasan Da\u{g}} 
\address{Management Information Systems Department, T. C. Kadir Has University, Istanbul, Turkey}
\cortext[cor1]{Corresponding author}

\begin{abstract}
Behavior of malware varies concerning the malware types, which affects the strategies of the system protection software. Many malware classification models, empowered by machine and/or deep learning, achieve superior accuracies for predicting malware types. Machine learning-based models need to do heavy feature engineering work, which affects the performance of the models greatly. On the other hand, deep learning-based models require less effort in feature engineering when compared to that of the machine learning-based models. However, traditional deep learning architectures’ components, such as max and average pooling, cause architecture to be more complex and the models to be more sensitive to data. The capsule network architectures, on the other hand, reduce the aforementioned complexities by eliminating the pooling components. Additionally, capsule network architectures based models are less sensitive to data, unlike the classical convolutional neural network architectures. This paper proposes an ensemble capsule network model based on the bootstrap aggregating technique. The proposed method is tested on two widely used, highly imbalanced datasets (Malimg and BIG2015), for which the-state-of-the-art results are well-known and can be used for comparison purposes. RCNF achieves the highest F-Score, which is $0.9820$, for the BIG2015 dataset and F-Score, which is $0.9661$, for the Malimg dataset. RCNF reaches the-state-of-the-art with fewer trainable parameters than other competitors.
\end{abstract}


\end{frontmatter}


\section{Introduction}
\label{S:1}
Malware type classification is as important as malware detection problem because system protection software makes their strategies concerning malware family types. Malware families have different behaviors and effects on a computer system. Each malware family uses different resources, files, ports, and other components of operating systems. For example, malware in the online banking systems aim to perform fraud, steal private information of users, and use different spreading behaviors \cite{etaher2015zeus, azab2016machine}. In addition to this, due to the trends in technology, new malware types occur almost daily. Thus, most of the computers, smartphones, and other digital systems are vulnerable to new malware. In this case, many zero-day attacks are performed \cite{alazab2013information}. The raising of the number of malware makes using the big data techniques crucial for malware analysis \cite{tang2017big}. 

Malware type classification is the most common problem in the cybersecurity domain, because strategies of protection systems vary with respect to malware family type. Malware type classification problem is broadly dealt with in three different ways:  static, dynamic, and image-based \cite{nataraj2011comparative, abijah2019intelligent, ni2018malware}. This paper focuses on the image-based malware family classification problem. However, malware family type classification is an imbalanced task, so this makes many models unsuccessful at predicting rare classes. To this end, two imbalanced datasets are used and the results are compared to other models in the literature.  

This paper proposes a new model named {\textbf R}andom {\textbf C}apsule {\textbf N}etwork {\textbf F}orest (RCNF) based on bootstrap aggregation (bagging) ensemble technique and capsule network (CapsNet) \cite{breiman1996bagging, sabour2017dynamic}. The main motive behind the proposed method is to reduce the variance of different CapsNet models (as weak learners) using bagging. In this perspective, the main contributions of this paper can be listed as follows:

\begin{itemize}
  \item The paper introduces the first application of CapsNet in the field of malware type classification. Although image-based malware classification is a broad research and application area, there is no research and application of CapsNet in the literature in our best knowledge.
  
  \item The paper uses the first ensemble model of CapsNet. The key idea of creating an ensemble of CapsNet is assuming a single CapsNet model as a weak classifier like a decision tree model. In this way, an ensemble model of CapsNet can be easily created using bootstrap aggregating. The assumption that CapsNet is a weak learner increases the performance of a single CapsNet for two different well-known malware datasets, which are highly imbalanced.  
  
  \item The proposed model uses simple architecture engineering instead of complex convolutional neural network architectures and domain-specific feature engineering techniques. In addition to this, CapsNet does not require to use transfer learning, and the model is easily trained from scratch. Because of that, the created network and its ensemble version have reasonably a lower the number of parameters.
  
  \item The proposed model is compared with the latest studies that use deep neural networks for image-based malware classification tasks. For a fair comparison,  especially,  the last studies using the Malimg and the BIG2015 datasets are chosen and compared with the proposed method.
  
\end{itemize}
Image-based malware classification is a broad research and application area. At the same time, deep learning drives the computer vision and image processing researches. Many deep convolutional neural networks have proven their success in image processing. CapsNet is the most important deep convolutional neural architecture that removes pooling to avoid losing the spatial features of images. This is the power of CapsNet comparing to classical CNNs. Therefore, the number of applications of CapsNet is increasing in image processing. The main motivation of this paper is to design a simple and accurate classifier for imbalanced malware type classification problem using bagging \cite{breiman1996bagging, breiman2001random} and CapsNet architecture. This paper also presents a detailed comparison of the proposed model with the other models in the literature.

The paper is organized as follows. Section \ref{sec:relatedwork} presents a literature survey for CapsNet applications and previous malware analysis studies. The methodology of the paper is described in Section \ref{sec:methodology}, whereas Section \ref{sec:model} gives details of the inspiring model and the proposed model. In Section \ref{sec:res}, the test results are discussed and many comparisons with related works published in the last years are listed; and finally, Section \ref{sec:conc} provides the concluding remarks.

\section{Related Work}
\label{sec:relatedwork}

There are many different ways to represent malware files for machine learning-based identification. One of them is to extract features from the application programming interface (API) calls of malware. For example, Alazab \cite{alazab2015profiling} proposed a framework to get features statically and dynamically from malware API calls. He used similarity mining and machine learning to profile and classify malware. He obtained a $0.966$ received-operating-curve (ROC) score in the malware dataset containing $66,703$ samples (Malign or Benign) with the k-nearest neighbor algorithm. Moreover, Azab et al. \cite{azab2014mining} focused on grouping malware in the same variants using hashing techniques for malware binaries. They used two different Zeus datasets. The first dataset contained 856 binaries, and the second dataset contained 22 binaries.  Each binary had SHA256 value. They achieved $0.999$ F-Score using the k-nearest neighbors and SDHASH.

The second efficient way to feed machine learning algorithms to classify malware files is image-based representation. In this work, we have focused on image-based malware type classification. For example, Nataraj et al. \cite{nataraj2011malware} converted malware files to greyscale images to represent malware. They extracted GIST features from malware images, and then they classified malware family types, using the euclidean k-nearest neighbors algorithm. They reached $0.98$ classification accuracy for the dataset that has $9339$ samples and $25$ malware families. Similarly, Kancherla et al. \cite{kancherla2013image} used image-based malware representation to feed the support vector machine to classify malware files malign or benign. They extracted three different features, such as intensity-based, wavelet-based, and Gabor based. Their dataset contained $15,000$ malign and $12,000$ benign samples. They attained $0.979$ ROC score. These studies utilized traditional machine learning algorithms, such as k-nearest neighbors, and the support vector machines. These algorithms required to extract good features from images to classify malware types with a high performance. After an impactful success of a deep convolutional neural network (CNN) on the Imagenet dataset, a new era started in computer vision \cite{krizhevsky2012imagenet}. CNNs could classify images using raw pixel values without complex feature engineering methods. 

Image-based malware classification is a broad research area, which is affected by deep convolutional networks. In this regard, one of the most important applications of CNNs is the transfer learning, which is useful and successful for the balanced and relatively small size datasets \cite{nahmias2019trustsign}. Sang Ni et al. \cite{ni2018malware} created greyscale image files using SimHash bits of malware. They obtained $0.9926$ accuracy on the dataset containing $10,805$ samples, using a CNN classifier. There are many deep learning models to classify malware types, but we used some of them in the experiment part of the paper. 

CapsNet, as a new CNN structure, has been implemented in 2017 \cite{sabour2017dynamic}, especially in the health domain \cite{jimenez2018capsule}, CapsNet have many applications in the literature. For instance, Afshar et al. \cite{afshar2018brain} use CapsNet for brain tumor classification problems like classification of breast cancer histology images of \cite{iesmantas2018convolutional}. Mobiny et al. \cite{mobiny2018fast} create a fast CapsNet architecture for lung cancer diagnosis. Another important application area of CapsNet is object segmentation. LaLonde et al. \cite{lalonde2018capsules} use CapsNet for object segmentation. Traditional CNN structures, on the other hand,  are  used in {\textbf G}enerative {\textbf A}dversarial {\textbf N}etworks, GANs. CapsNet is very useful to make GANs better by removing the weakest point of these CNNs \cite{jaiswal2018capsulegan}. The snippet studies above show that CapsNet is a promising architecture against the standard CNN. In the literature, although there are many applications of CapsNets, there is a missing and important area. This area is a computer and information security. This gap can be seen easily in the pre-print version of a survey about CapsNets \cite{patrick2019capsule}.

Another crucial issue in malware classification is imbalanced datasets. Ebenuwa et al. \cite{ebenuwa2019variance} pointed imbalanced classification problem in binary classification. They inspected three different techniques, such as sampling-based, algorithm modifications, and cost-sensitive approaches. They proposed variance ranking feature selection techniques to get better results in imbalanced datasets for binary classification problems.

To this end, this paper aims to develop a malware classification model based on an ensemble of CapsNet architecture for imbalanced malware datasets, which is the first application of CapsNets in the malware classification domain.

\section{Methodology}
\label{sec:methodology}
\subsection{Malware Datasets}
\label{sec:datasets}

There are many open research issues in malware classification. These issues can be listed such as class imbalance, concept drift, adversarial learning, interpretability (explainability) of the models, and public benchmarks \cite{gibert2020rise}. In this paper, our model called RCNF focuses on the class imbalance issue. Thus, the base CapsNet and the proposed RCNF models have been tested on two very well-known malware datasets called Malimg and Microsoft Malware 2015 (BIG2015). These datasets are highly imbalanced in terms of the number of classes. This section describes these datasets.

\subsubsection{Malimg}
\label{subsec:malimg}
Nataraj et al. introduced a new malware family type classification approach based on visual analysis, converted binaries into greyscale images and they published these images as a new malware dataset called Malimg \cite{nataraj2011malware}. This dataset has $9339$ samples and $25$ different classes. Table \ref{tbl:malimg_sample_dist} presents the number of samples for each malware family. This distribution shows that the dataset is highly imbalanced.

\begin{table}[!h]
\centering
\caption{Sample Distribution for each Malware Family.}
\begin{tabular}{|c|c|c|}
\hline
No. &   Family Name    & Number of Samples \\ \hline
$1$ &    Allaple.L      & $1591$              \\ \hline
$2$ &    Allaple.A      & $2949$              \\ \hline
$3$ &    Yuner.A        & $800$               \\ \hline
$4$ &   Lolyda.AA $1$    & $213$               \\ \hline
$5$ &    Lolyda.AA $2$    & $184$               \\ \hline
$6$ &    Lolyda.AA $3$    & $123$               \\ \hline
$7$ &    C2Lop.P        & $146$               \\ \hline
$8$ &    C2Lop.gen!g    & $200$               \\ \hline
$9$ &    Instantaccess  & $431$               \\ \hline
$10$ &    Swizzot.gen!I  & $132$               \\ \hline
$11$ &    Swizzor.gen!E  & $128$               \\ \hline
$12$ &    VB.AT          & $408$               \\ \hline
$13$ &    Fakerean       & $381$               \\ \hline
$14$ &    Alueron.gen!J  & $198$               \\ \hline
$15$ &    Malex.gen!J    & $136$               \\ \hline
$16$ &    Lolyda.AT      & $159$               \\ \hline
$17$ &    Adialer.C      & $125$               \\ \hline
$18$ &    Wintrim.BX     & $97$                \\ \hline
$19$ &    Dialplatform.B & $177$               \\ \hline
$20$ &    Dontovo.A      & $162$               \\ \hline
$21$ &    Obfuscator.AD  & $142$               \\ \hline
$22$ &    Agent.FYI      & $116$               \\ \hline
$23$ &    Autorun.K      & $106$               \\ \hline
$24$ &    Rbot!gen       & $158$               \\ \hline
$25$ &   Skintrim.N     & $80$                \\ \hline
\end{tabular}
\label{tbl:malimg_sample_dist}
\end{table}

\begin{figure}
\begin{subfigure}{.5\textwidth}
  \centering
  \includegraphics[width=.6\linewidth]{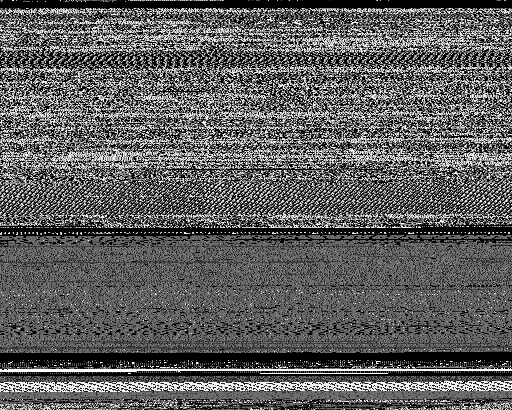}
  \caption{An image example from Family Adialer.}
  \label{fig:adlalerc}
\end{subfigure}%
\begin{subfigure}{.5\textwidth}
  \centering
  \includegraphics[width=.6\linewidth]{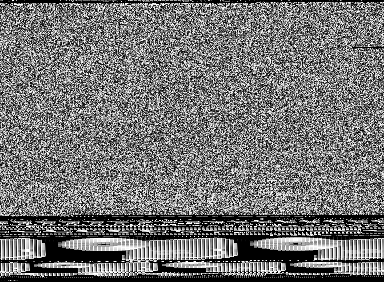}
  \caption{An image example from Family Fakerean}
  \label{fig:fakerean}
\end{subfigure}
\caption{Malware image samples obtained from byte files using the algorithm described in \cite{nataraj2011malware}.}
\label{fig:malimg_samples}
\end{figure}

Fig. \ref{fig:malimg_samples} shows the malware images created from the byte files. All images are single-channel and are resized to $224 \times 224$ for CapsNet architecture. This size is the largest value that can be processed in our computer system. 

\subsubsection{Microsoft Malware 2015 (BIG2015)}
\label{subsec:microsoft2015}
\begin{figure}[h!]
\begin{center}
  \includegraphics[width=0.5\linewidth]{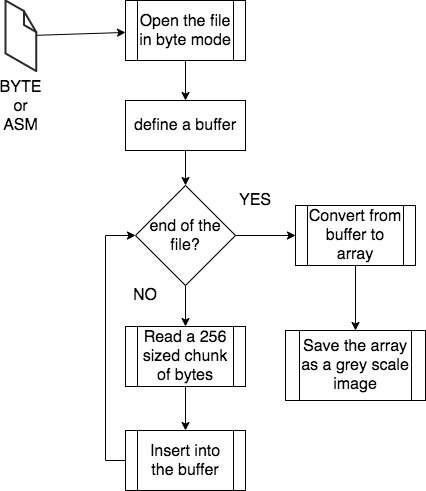}
  \caption{Flowchart of the process of a flat-file to greyscale image for the BIG2015 dataset.}
\label{fig:file2image}
\end{center}
\end{figure}

BIG2015 dataset has been released as a Kaggle competition \cite{ronen2018microsoft, Microsof31:online}. Table \ref{tbl:distribution_ms2015} presents the sample distribution for each malware family in BIG\-2015 dataset. The distribution shows that the dataset is highly imbalanced; and \emph{Simda} is the toughest malware family to be predicted for the dataset. The dataset contains $10868$ BYTE (bytes) files and $10868$ ASM (assembly code) files and $9$ different malware family types. BIG2015, unlike the Malimg dataset, contains raw files. Thus, a file from the BIG2015 dataset is opened in the byte mode and then the file is read by $256$ sized chunks till the end of the file. Finally, the buffer is converted to array and the array is saved as a greyscale image into the file system. All processes are described in Fig. \ref{fig:file2image}. This method is the most common way to convert from malware files to images \cite{nataraj2011malware, venkatraman2018use}.

\begin{table}[h]
\centering
\caption{Number of Samples for Each Malware Family in BIG2015 Dataset.}
\begin{tabular}{|c|c|c|}
\hline
No. & Family Name    & Number of Image \\ \hline
$1$   & Ramnit         & $1541$            \\ \hline
$2$   & Lollipop       & $2478$            \\ \hline
$3$   & Kelihos\_ver3  & $2942$            \\ \hline
$4$   & Vundo          & $475$             \\ \hline
$5$   & Simda          & $42$              \\ \hline
$6$   & Tracur         & $751$             \\ \hline
$7$   & Kelihos\_ver1  & $398$             \\ \hline
$8$   & Obfuscator.ACY & $1228$            \\ \hline
$9$   & Gatak          & $1013$            \\ \hline
\end{tabular}
\label{tbl:distribution_ms2015}
\end{table}

\begin{figure}
\begin{subfigure}{.5\textwidth}
  \centering
  \includegraphics[width=.6\linewidth]{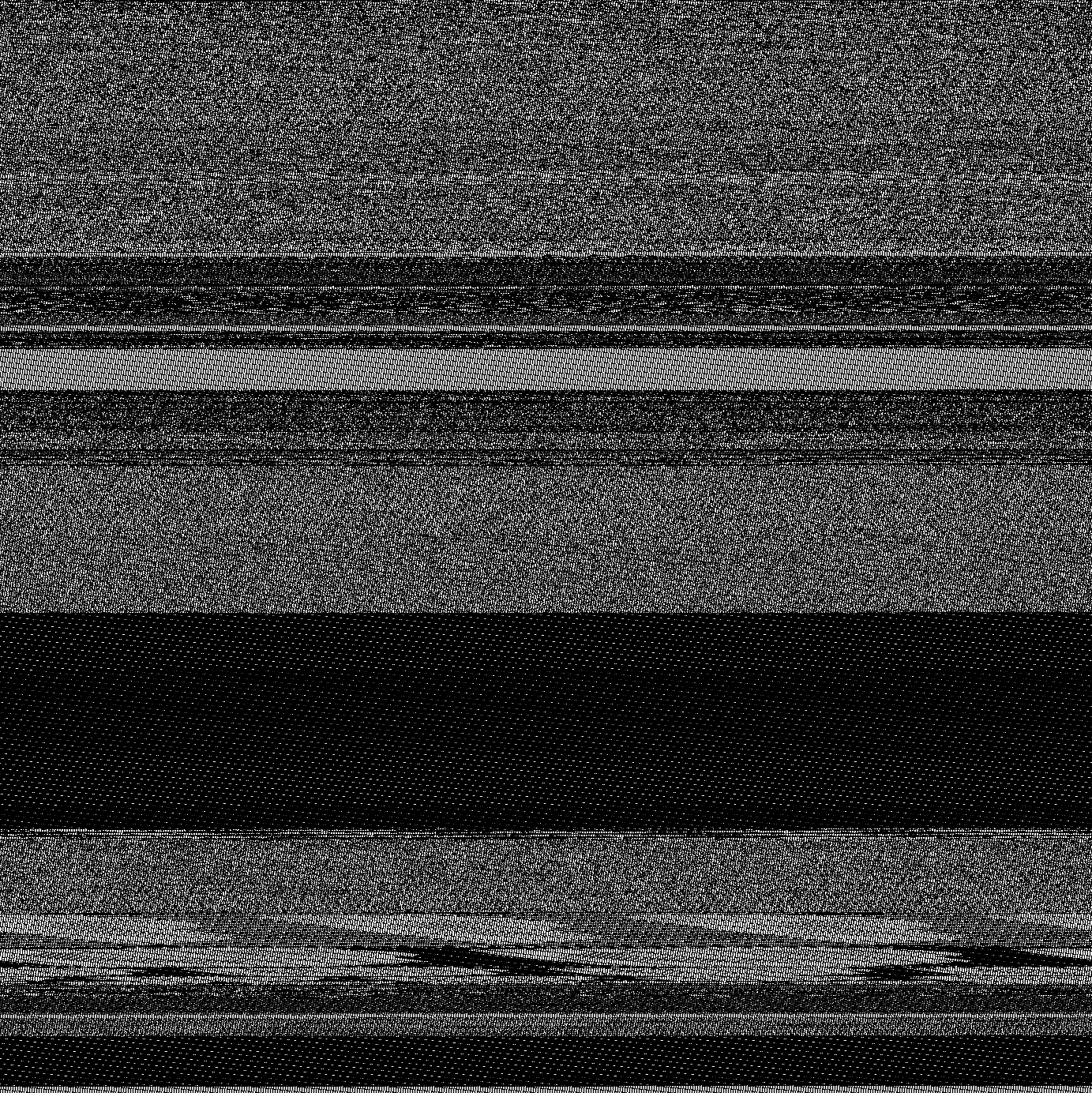}
  \caption{An image example obtained from BYTE File \\ in Ramnit Family}
  \label{fig:sfig1}
\end{subfigure}%
\begin{subfigure}{.5\textwidth}
  \centering
  \includegraphics[width=.6\linewidth]{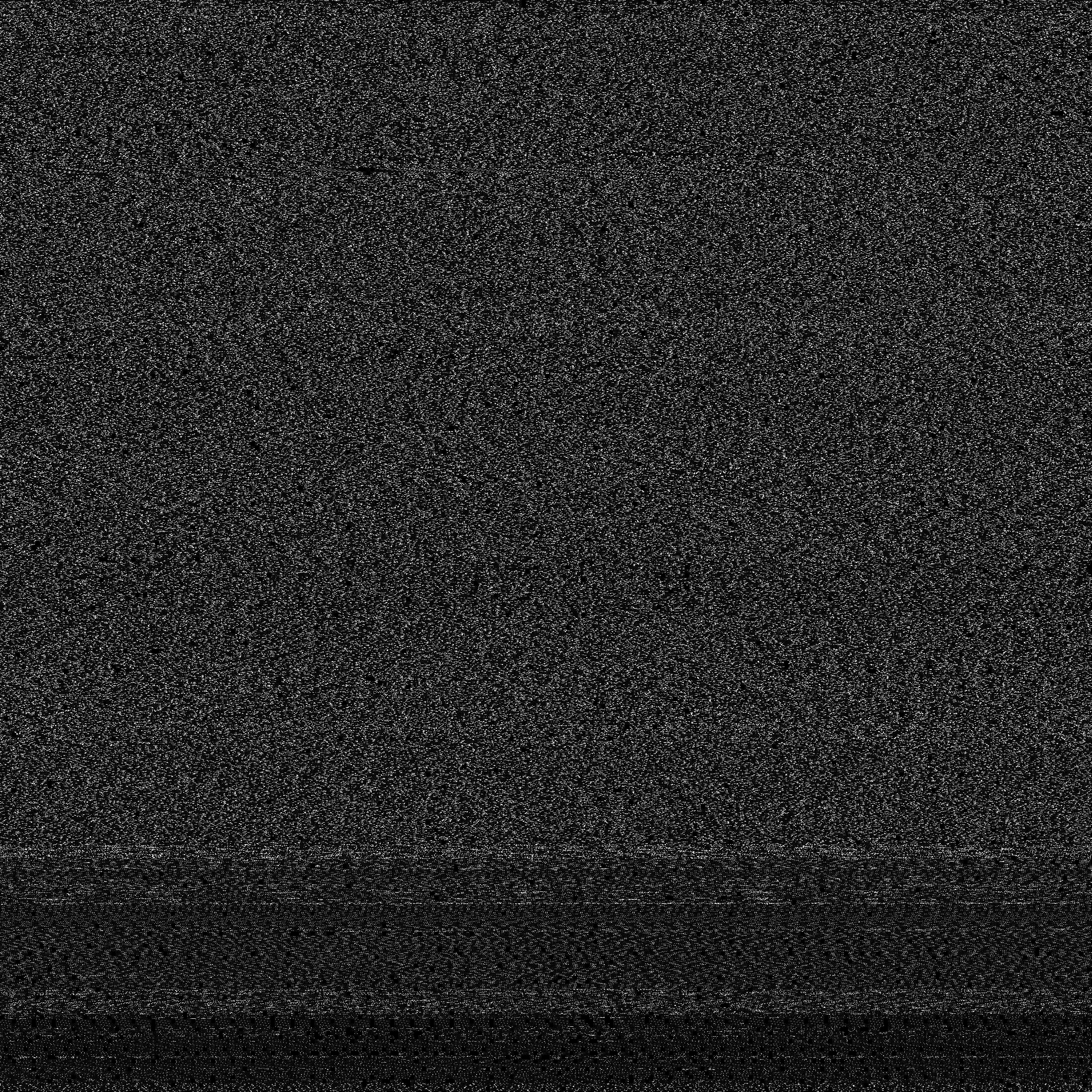}
  \caption{An image example obtained from ASM File \\ in Ramnit Family}
  \label{fig:sfig2}
\end{subfigure}
\caption{BIG2015 image samples from BYTE and ASM files using the algorithm described in \cite{nataraj2011malware}.}
\label{fig:ms2015_samples}
\end{figure}

Fig. \ref{fig:ms2015_samples} depicts image representations created from the BYTE and ASM files of the same malware sample in \emph{Ramnit} malware family. All images are single channel. All images are resized $112 \times 112$ for our CapsNet architecture, because the architecture uses both BYTE and ASM image representations at the same time.

\section{Model}
\label{sec:model}
In this section, general capsule networks, base CapsNet architecture for Malimg and base CapsNet architecture for BIG2015 are described. CapsNet architectures are different for both Malimg and BIG2015 Dataset.
\subsection{Capsule Networks}
\label{subsec:capsnet}
Capsule networks are special architectures of convolutional neural networks aiming to minimize information loss because of max pooling \cite{sabour2017dynamic}. This method is the weakest point for preserving spatial information \cite{iesmantas2018convolutional}. A CapsNet contains capsules similar to autoencoders \cite{krizhevsky2011using, sabour2017dynamic}. Each capsule learns how to represent an instance for a given class. Therefore, each capsule creates a fixed-length feature vector to be input for a classifier layer without using max pooling layers in its internal structure. In this way, this capsule structure aims to preserve texture and spatial information with minimum loss.

Sabor et al. propose an efficient method to train CapsNet architectures \cite{sabour2017dynamic}. This method is called a dynamic routing algorithm, which uses a new non-linear activation function called squashing shown in  (\ref{eqn:squashing}). This equation emphasizes that short vectors are shrunk to almost zero and long vectors are shrunk to $1$ \cite{sabour2017dynamic}. In this equation, $v_{i}$ is the output of $i^{th}$ capsule and $s_{i}$ shows the total input of this capsule.

\begin{equation}
\label{eqn:squashing}
    v_{i} = \frac{\norm{s_{i}}^2}{1 + \norm{s_{i}}^2}  \: \frac{s_{i}}{\norm{s_{i}}} 
\end{equation}

\begin{figure}[h!]
\begin{center}
  \includegraphics[width=\linewidth]{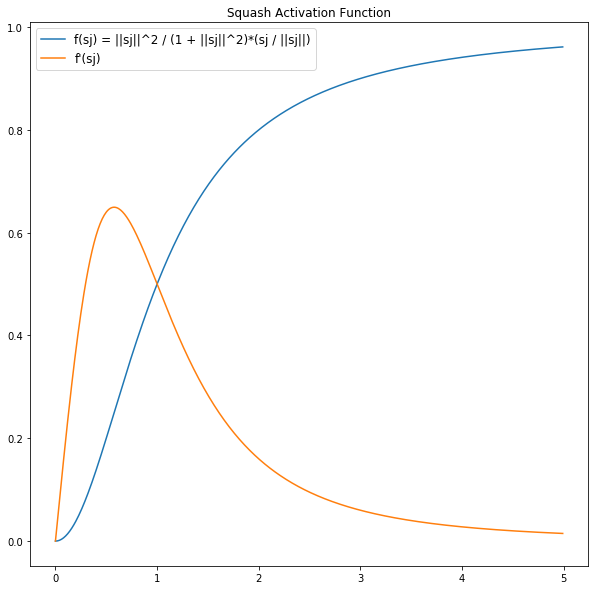}
  \caption{Squashing activation function and its derivation in 2-D plane.}
\label{fig:squash_act}
\end{center}
\end{figure}

Visualizing the squash activation function described in (\ref{eqn:squashing}) is hard because its input is a high dimensional vector. If the activation function can be thought of as a single variable function, as described in \cite{marchisio2019capsacc}, then the behavior of the function and its derivative can be visualized, as in Fig. \ref{fig:squash_act}.

\begin{figure}[h!]
\begin{center}
  \includegraphics[width=\linewidth]{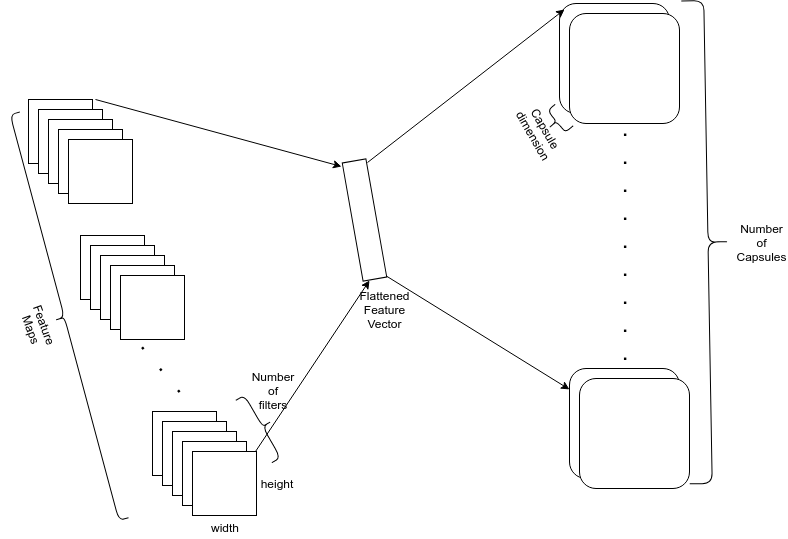}
  \caption{Basic CapsNet architecture.}
\label{fig:basic_capsnet}
\end{center}
\end{figure}

A basic CapsNet architecture contains two parts: the standard convolution blocks and the capsule layer as shown in Fig. \ref{fig:basic_capsnet}. A convolution block is made from a combination of convolution filters and ReLU activation function. At the end of the convolution block, obtained feature maps are reshaped and projected to $d-dimensional$ vector representation. This representation feeds each capsule in the capsule layer. Each capsule learns how to represent and reconstruct a given sample like an autoencoder \cite{krizhevsky2011using} architecture. In order to learn how to reconstruct a malware sample, the capsule network will minimize reconstruction error in (\ref{eqn:reconloss}), where $x_{c} \in R^{dxd}$ is the real sample in the capsule $c$ and $\hat{x}_{c} \in R^{dxd}$ is the reconstructed sample by the same capsule $c$. These representations are used to calculate the class probabilities for the classification task.

\begin{equation}
\label{eqn:reconloss}
    \ell_{r} = (x_{c} - \hat{x}_{c})^2
\end{equation}

Margin loss function is used for CapsNet. This function is similar to hinge loss \cite{rosasco2004loss}. (\ref{eqn:marginloss}) defines the margin loss $L_{c}$ for capsule $c$, 

\begin{equation}
\label{eqn:marginloss}
    \ell_{m} = y_{c} \times (max(0, m - \hat{y}_{c}))^{2} + \lambda \times (1 - y_{c}) \times (max(0,  \hat{y}_{c} - (1 - m)))^{2} 
\end{equation}
where $m = 0.9$, $\lambda = 0.5$, $y_{c}$ denotes actual class and $\hat{y}_{c}$ represents the current prediction.

\begin{equation}
\label{eqn:loss}
    L_{c} = \ell_{m} + 0.0005 \times \ell_{r}
\end{equation}

\begin{equation}
\label{eqn:loss}
    L = \frac{1}{N} \sum_{n=1}^{N} L_{c}
\end{equation}

The mean of $L_{c}$ for each capsule gives the total loss in (\ref{eqn:loss}), where $L_{c}$ is sum of margin loss $\ell_{m}$ (as described in (\ref{eqn:marginloss})) and reconstruction loss $\ell_{r}$ (as described in (\ref{eqn:reconloss})). However, reconstruction loss is multiplied by 0.0005 to avoid suppressing the margin loss \cite{sabour2017dynamic}. In order to minimize loss $L$, one can use the most applicable optimizer algorithm for CapsNet by Adam \cite{kingma2014adam, sabour2017dynamic}. We have observed that CapsNet cannot converge to minimum loss value with optimizers other than that of Adam. This is obviously an open issue for CapsNet studies in the future. 

In image-based malware family type classification problem, there are no complex patterns that are easily detected by classical convolutional neural networks. For this reason, the predictive model must recognize the pattern of pixel distribution of the image-based malware sample. On the other hand, CapsNet can learn pixel density distribution of each malware family. Thus, a CapsNet model can be easily trained from scratch for this problem, unlike CNNs. This is the most important advantage of using CapsNet as a base classifier in our proposed model.

Our main assumption is that CapsNet architecture will be able to successfully classify malware family types using raw pixel values obtained from malware binary and assembly files. In addition to the main assumption, this paper aims to increase CapsNet malware type classification architecture accuracy with the bagging ensemble method.

\subsection{Base Capsule Network Model for Malimg Dataset}
\label{subsec:basecapsmalimg}
\begin{figure}[h!]
\begin{center}
\includegraphics[width=140mm,height=40mm]{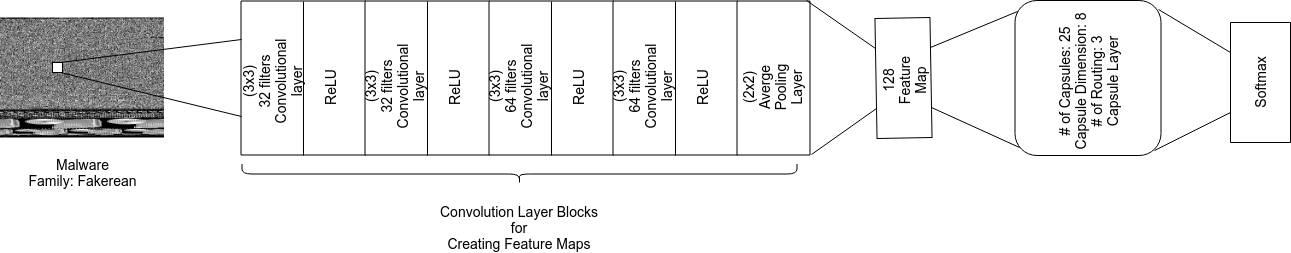}
  \caption{CapsNet architecture for Malimg dataset.}
\label{fig:malimg_capsnet}
\end{center}
\end{figure}

Before creating an ensemble CapsNet model, the base CapsNet estimator must be built. This architecture depends on the dataset. Thus, base CapsNet estimator architecture has a single convolution line, as shown in Fig. \ref{fig:malimg_capsnet}. The convolutional line contains two sequential blocks; and each block contains two sequential convolutions and ReLU layers. The first two convolutional layers have $3\times3$ kernels and $32$ filters. The second two convolutional layers have $3\times3$ kernels and $64$ filters. Feature maps are reshaped to $128$ feature vectors. At the end of the reshape step, there is a capsule layer containing $25$ capsules; the dimension of each capsule is $8$; and the routing iteration is $3$ of the capsule layer. This is the optimal CapsNet architecture for the Malimg dataset depending on our experiments.

\subsection{Base Capsule Network Model for BIG2015 Dataset}
\label{subsec:basecapsms2015}
\begin{figure}[h!]
\begin{center}
\includegraphics[width=140mm,height=60mm]{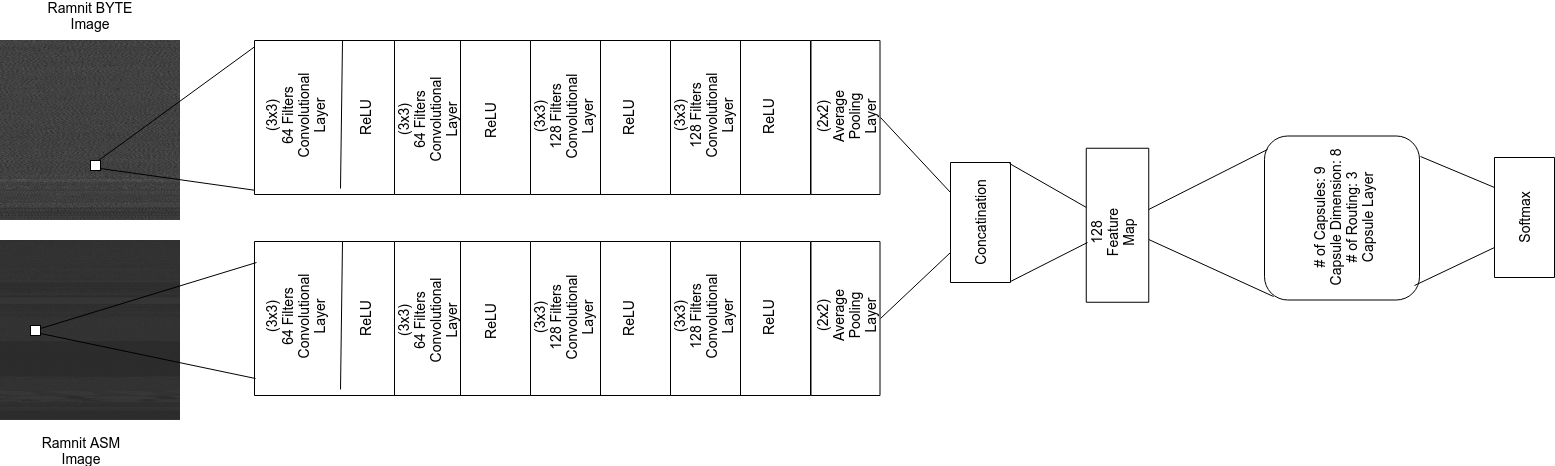}
  \caption{CapsNet architecture for BIG2015 dataset.}
\label{fig:big2015_capsnet}
\end{center}
\end{figure}

The BIG2015 dataset has two different files for each sample. One of them is a binary file; and the other is an assembly file. Thus, it is possible to design a CapsNet, which can be fed by two different image inputs at the same time. Fig. \ref{fig:big2015_capsnet} shows a CapsNet architecture, which has two exactly identical convolution lines. In this architecture, the first two sequential layers contain $3\times3$ kernels and $64$ filters. The second two sequential layers contain $3\times3$ kernels and $128$ filters.  

Features extracted from the ASM and BYTE images are concatenated; and the final feature vector is reshaped to a vector with length $128$. For the next level, as an input, this feature vector feeds to a capsule layer containing $9$ capsules. In this layer, the dimension of each capsule is $8$;  and the routing iteration is $3$. This hyper-parameter set is optimal for the base CapsNet estimator for the BIG2015 dataset. 

\subsection{The Proposed Random CapsNet Forest Model for Inbalanced Datasets}
\label{subsec:proposedmodel}
Random CapsNet Forest (RCNF) is an ensemble model, which is inspired by the Random Forest algorithm \cite{breiman2001random}. The basic idea behind RCNF is to assume identical CapsNet models as weak learners create different training sets for each model from the original training set using the bootstrap resampling technique, as shown in Algorithm \ref{alg:training}. The training algorithm is a variant of bootstrap aggregating (also known as bagging) \cite{breiman1996bagging} for CapsNet model; and bagging reduces the variance of the model while increasing robustness of the model \cite{breiman1996bias}. In this paper, bagging is preferred to create an ensemble of CapsNet instead of boosting \cite{freund1996experiments}, because it is shown that boosting tends to overfit \cite{quinlan1996bagging}. During the training phase, each epoch updates the weights of the CapsNet. Therefore, the weight of the best model at the end of each epoch is saved according to the validation score to increase model performance and consistency against random weight initialization of the CapsNet.

\begin{algorithm}[!h]
\caption{Random CapsNet Forest Training Algorithm}\label{alg:training}
\begin{algorithmic}[1]
\Procedure{train}{$base\_model, n\_estimators, trainset, valset, epochs$}
\For{$i \gets 1, n\_estimators$}
\State $bs\_trainset \gets resample(trainset, replacement=True)$
\For{$e \gets 1, epochs$}
\State $base\_model.fit(bs\_trainset)$
\State $val\_score \gets get\_accuracy(base\_model, valset)$
\If{$is\_best\_score(val\_score) == True$}
\State $save\_weights(base\_model)$
\EndIf
\EndFor
\EndFor
\EndProcedure
\end{algorithmic}
\end{algorithm}

\begin{algorithm}[!h]
\caption{Random CapsNet Forest Prediction Algorithm}\label{alg:prediction}
\begin{algorithmic}[1]
\Procedure{predict}{$n\_estimators, testset, numclasses$}\Comment{Average Ensembling}
\State $total\_preds \gets zeros\_like(testset.shape[0], numclasses)$
\For{$i \gets 1, n\_estimators$}
\State $model_{i} \gets load\_model\_weights(i)$
\State $total\_preds += model_{i}.predict(testset)$
\EndFor
\State $preds \gets total\_preds / n\_estimators$
\State \textbf{return} argmax($preds$)\Comment{The final predictions of CapsNet models}
\EndProcedure
\end{algorithmic}
\end{algorithm}

The prediction method is described in Algorithm \ref{alg:prediction}. Each weight of CapsNet model is loaded; and test samples are predicted by the model. Cumulative predicted probabilities are added onto $total\_preds$ variable; and this step is known as average ensembling step. At the end of the estimation loop, the index of the highest probabilities is assigned as a predicted class.  

There are several limitations to the RCNF model. The first limitation is the number estimators in the RCNF model. In this implementation, an RCNF model can contain up to $10$ CapsNets because of an increasing number of trainable parameters. The second limitation is the training time. Training of an RCNF with $10$ CapsNets for the BIG2015 dataset takes five hours ($100$ epochs for each CapsNet). This training time of the RCNF with $5$ CapsNets model for the Malimg dataset is decreasing ($100$ epochs for each CapsNet). On the other hand, the RCNF can be easily parallelized to increase efficiency in the training phase. Each CapsNet can be trained on multiple GPUs. We will develop a distributed multi-GPU version of the RCNF as a future work. 

We implemented the RCNF using Tensorflow (version $1.5$) \cite{abadi2016tensorflow} and Keras \cite{chollet2015keras}, Sklearn \cite{scikit-learn}, Numpy \cite{oliphant2006guide} and Pandas \cite{mckinney2010data}. All scripts were written in Python3. The configuration of the computer used in this study was $12$GB GPU (GeForce GTX $1080$ Ti) and Intel Core i9-9900K processor with $64$ GB main memory for testing.

\section{Experiment and Results}
\label{sec:res}
CapsNet and RCNF ensemble models are tested on two different datasets called Malimg and BIG2015. The Malimg dataset has been divided into three parts: training, validation, and test sets. The training set has $7004$ samples, the validation set has $1167$ samples, and the test set has $1167$ samples. BIG2015 has also been divided into three parts like the Malimg dataset. In the experiments of CapsNet and RCNF ensemble model for the BIG2015 dataset, the training set has $8151$ samples, the validation set has $1359$ samples and the test set has $1358$ samples.  
The first experiment is made to obtain performance results of single base CapsNet estimators for each dataset. The second experiment is about the performance of the RCNF model. Model evaluation has been done in terms of accuracy, F-Score, and the number of parameters of deep neural nets. These performance metrics are defined as follows:  
\begin{equation}
\label{eqn:accuracy}
    accuracy = \frac{(TP + TN)}{(TP + TN + FP + FN)} 
\end{equation}
\begin{equation}
\label{eqn:f1score}
    F\text{-}Score = \frac{2 \times TP}{2 \times TP + FN + FP}
\end{equation}
where true positive (TP) and false positive (FP) are the numbers of instances correctly and wrongly classified as positive respectively. True negative (TN) and false negative (FN) are the number of instances correctly and wrongly classified as negative respectively. Accuracy is the ratio of the number of true predictions to all instances in the set as shown in  (\ref{eqn:accuracy}). F-Score is shown as the set (\ref{eqn:f1score}) in terms of true positives, false negatives, and false positives. Accuracy is not a correct performance metric for imbalanced datasets. On the other hand, papers compared in this work use accuracy and F-Score performance metrics to measure the success of their models. Thus, this paper gives the experiment results in terms of accuracy and the F-Score. Our main goal is that an ensemble of the CapsNet can reduce the number of FN and FP in (\ref{eqn:f1score}).

\begin{figure}[htbp]
 \centering
\begin{subfigure}{0.5\textwidth}
\includegraphics[width=85mm,scale=0.5]{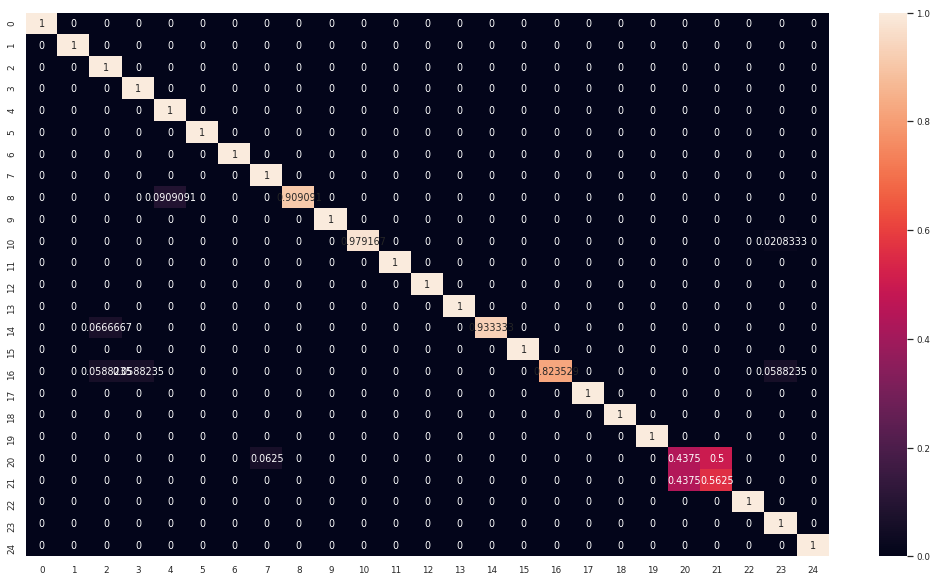}
\caption{Malimg Test Set}
\label{fig:single_malimg}
\end{subfigure}
\begin{subfigure}{0.5\textwidth}
\includegraphics[width=85mm, scale=0.5]{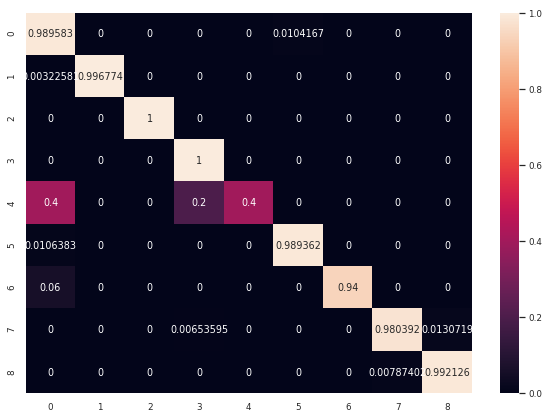}
\caption{BIG2015 Test Set}
\label{fig:single_big2015}
\end{subfigure}
 
\caption{Confusion Matrices of single CapsNet Model for each test set.}
\label{fig:single_capsnet_cm}
\end{figure}

Fig. \ref{fig:single_capsnet_cm} shows confusion matrices for each test part of both datasets. Each confusion matrix (Fig. \ref{fig:single_malimg} and \ref{fig:single_big2015}) implies that a model containing single CapsNet incorrectly predicts rare malware families in both datasets.
 
\begin{figure}[h!]
\begin{center}
  \includegraphics[width=\linewidth]{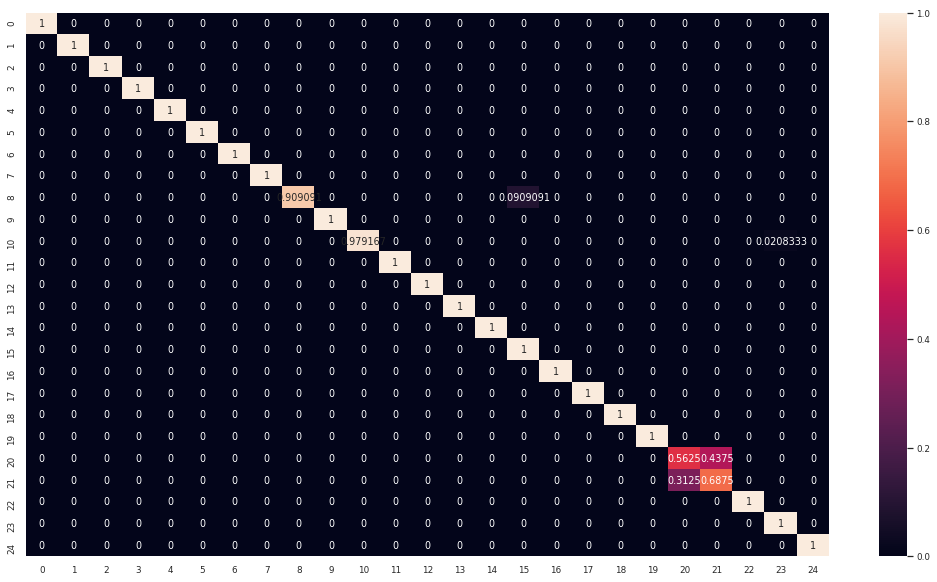}
  \caption{Confusion Matrix of 5-RCNF for Malimg test set.}
\label{fig:malimg_capsnet_cm}
\end{center}
\end{figure}

Fig. \ref{fig:malimg_capsnet_cm} is the confusion matrix of RCNF containing $5$ base CapsNet models. This confusion matrix shows the prediction accuracy of the model for each malware family type in the Malimg test set. Class $8$, $10$, $20$, and $21$ have been predicted wrongly by the RCNF model. On the other hand, the model has been very successful at correctly predicting other malware types in the test set. This confusion matrix also shows that RCNF is successful at correctly predicting rare malware types in the Malimg test set. 

In the second experiment, RCNF is tested on the BIG2015 dataset. Fig. \ref{fig:malimg_capsnet_cm} shows the prediction results of RCNF containing $10$ base CapsNet for BIG2015 dataset. Class $4$ is the rarest malware type in the whole dataset. Training, validation, and test sets are stratified, so the class distribution is preserved for each partition. This result shows that RCNF can predict the rarest malware type pretty well. Class $0$, $1$, $2$, and $6$ are predicted perfectly by RCNF. If the performance of RCNF is compared with the performance of a single CapsNet model, it is easily seen that RCNF is better than a single CapsNet at predicting rare malware families for imbalanced datasets. 

\begin{figure}[h!]
\begin{center}
  \includegraphics[width=\linewidth]{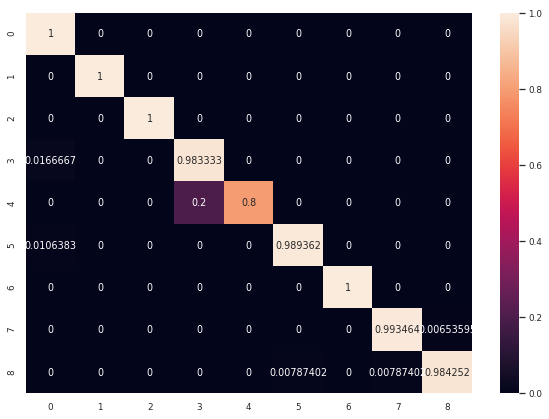}
  \caption{Confusion Matrix of 10-RCNF for BIG2015 test set.}
\label{fig:big2015_capsnet_cm}
\end{center}
\end{figure}

\begin{table}[h!]
\centering
\caption{Comparison RCNF and other methods for Malimg test set performance.}
\begin{tabular}{|c|c|c|c|}
\hline
Model                                                                   & Number of Parameters & F-Score        & Accuracy        \\ \hline
Yue \cite{yue2017imbalanced}                                                                    & $20$M                  & -               & $0.9863$          \\ \hline
Cui et al. \cite{cui2018detection}                                                             & -                    & $0.9455$          & $0.9450$          \\ \hline
Venkatraman et al. \cite{venkatraman2019hybrid}                                                             & $212,885$                    & $0.916$          & $0.963$          \\ \hline
Vasan et al. \cite{vasan2020imcfn}                                                             & $134$M                    & $0.9820$          & $0.9827$          \\ \hline
Vasan et al. \cite{vasan2020image}                                                             & $157$M                    & $0.9948$          & $0.9950$          \\ \hline
\textbf{\begin{tabular}[c]{@{}c@{}}CapsNet\\ for\\ Malimg\end{tabular}} & \textbf{$90,592$}       & \textbf{$0.9658$} & \textbf{$0.9863$} \\ \hline
\textbf{\begin{tabular}[c]{@{}c@{}}RCNF\\ for\\ Malimg\end{tabular}}    & \textbf{$5 \times 90,592$}   & \textbf{$0.9661$} & \textbf{$0.9872$} \\ \hline
\end{tabular}
\label{tbl:malimg_comp}
\end{table}

\begin{table}[h!]
\centering
\caption{Comparison RCNF and other methods for BIG2015 test set performance.}
\begin{tabular}{|c|c|c|c|}
\hline
Model                                                                      & Number of Parameters & F-Score        & Accuracy        \\ \hline
Venkatraman et al. \cite{venkatraman2019hybrid}                                                                 & $212,885$                    & $0.725$               & -            \\ \hline
Cao et al. \cite{cao2018efficient}                                                                 & -                    & -               & $0.95$            \\ \hline
Gibert et al. \cite{gibert2018end}                                                              & -                    & $0.9813$          & $0.9894$          \\ \hline
Kreuk et al. \cite{kreuk2018deceiving}                                                               & -                    & -               & $0.9921$          \\ \hline
Le et al. \cite{le2018deep}                                                                  & $268,949$               & $0.9605$          & $0.9820$          \\ \hline
Chen \cite{chen2018deep}                                                                       & -                    & -               & $0.9925$          \\ \hline
Jung et al. \cite{jung2018malware}                                                               & $148,489$               & -               & $0.99$            \\ \hline
Abijah et al. \cite{abijah2019intelligent}                                                             & -                    & -               & $0.9914$          \\ \hline
Zhao et al. \cite{zhao2019maldeep}                                                               & -                    & -               & $0.929$           \\ \hline
Khan et al. \cite{khan2018analysis}                                                               & -                    & -               & $0.8836$          \\ \hline
Safa et al. \cite{safa2019benchmarking}                                                               & -                    & -               & $0.9931$          \\ \hline
Kebede et al. \cite{kebede2017classification}                                                             & -                    & -               & $0.9915$          \\ \hline
Kim et al. \cite{kim2018classifying}                                                            & -                    & -               & $0.9266$          \\ \hline
Kim et al. \cite{kim2018detecting}                                                                 & -                    & $0.8936$          & $0.9697$          \\ \hline
Yan et al. \cite{yan2018detecting}                                                                 & -                    & -               & $0.9936$          \\ \hline
Naeem et al. \cite{naeem2019identification}                                                                 & -                    & $0.971$               & $0.9840$          \\ \hline
Jang et al. \cite{jang2020fasttext}                                                                 & -                    & -               & $0.9965$          \\ \hline
\textbf{\begin{tabular}[c]{@{}c@{}}CapsNet\\ for\\ BIG2015\end{tabular}} & \textbf{$527,232$}      & \textbf{$0.9779$} & \textbf{$0.9926$} \\ \hline
\textbf{\begin{tabular}[c]{@{}c@{}}RCNF\\ for\\ BIG2015\end{tabular}}    & \textbf{$10 \times 527,232$} & \textbf{$0.9820$} & \textbf{$0.9956$} \\ \hline
\end{tabular}
\label{tbl:big2015_comp}
\end{table}

Table \ref{tbl:malimg_comp} shows the test performance of the proposed models and others for the Malimg dataset. Yue \cite{yue2017imbalanced} uses a weighted loss function to handle the imbalanced class distribution problem in the Malimg dataset; and also uses the transfer learning \cite{yosinski2014transferable} method to classify malware family types. Due to using transfer learning, the architecture has 20M parameters; and the model is so huge. Cui et al. \cite{cui2018detection} use classical machine learning methods such as K-Nearest Neighbor and support vector machines. They have trained these algorithms using GIST and GLCM features, which are feature engineering methods for images; and they have applied resampling to the dataset to solve imbalanced dataset problems. RCNF does not use a weighted loss function or any sampling method to overcome the imbalanced dataset problem. Our results are higher than these two methods; and the results also show that CapsNet and RCNF do not require any method for extra feature engineering in the Malimg dataset. A single CapsNet architecture for the Malimg dataset has $90,592$ trainable parameters and RCNF has $452,960$ trainable parameters, so our proposed methods are reasonably smaller than Yue's method. Venkatraman et al. \cite{venkatraman2019hybrid} propose two different models called CNN BiLSTM and CNN BiGRU with two variants of these models, which are called cost-sensitive and cost-insensitive. When CNN BiGRU reaches its own highest F-Score and accuracy on the Malimg dataset, its number of the trainable parameters is greater than RCNF and its scores are lower than RCNF. In other words, RCNF reaches the-state-of-the-art scores in terms of F-Score and accuracy with a lower parameter size.

IMCFN \cite{vasan2020imcfn} is an important deep convolutional network for the Malimg dataset. When the network is trained from scratch without any data augmentation, it outperforms our RCNF implementation. IMCFN also has $126,727,705$ trainable parameters. When it is compared to RCNF, IMCFN is a huge convolutional network. On the other hand, RCNF is more accurate than IMCFN, but it has a lower F-Score than IMCFN. The most important advantage of RCNF is using a lower number of parameters, and this makes RCNF trainable in GPUs with low memory.

IMCEC \cite{vasan2020image} is one of the most successful an ensemble model of CNN architectures for the MALIMG in terms of accuracy and F-Score. However, IMCEC is more complex than RCNF. IMCEC uses two different CNN architectures VGG16 \cite{simonyan2014very} and ResNet-50 \cite{he2016deep} to use transfer learning. Although IMCEC is more accurate than RCNF in terms of accuracy and F-Score, IMCEC has a total of $157$M parameters because of using these CNN networks. Unlike IMCEC, RCNF has $452,960$ trainable parameters and it is reasonably accurate as much as IMCEC.

Table \ref{tbl:big2015_comp} compares the test performance of the proposed models and others for the BIG2015 dataset. Venkatraman et al. \cite{venkatraman2019hybrid} also test the CNN BiGRU model on the BIG2015 dataset. In this case, the performance of the CNN BiGRU model is lower than RCNF, but its number of trainable parameters are reasonably lower than RCNF. Chen \cite{chen2018deep} and Khan et al. \cite{khan2018analysis} use transfer learning architectures for the dataset. Test results of our proposed methods are better than those two models, but our results are very close to Chen \cite{chen2018deep} in terms of accuracy. Our proposed models are better than Gibert et al. \cite{gibert2018end} in terms of accuracy, but the F-Score of RCNF is very close to this model. Our proposed models are better than models of Cao et al. \cite{cao2018efficient}, Zhao et al. \cite{zhao2019maldeep}, Kim et al. \cite{kim2018classifying}, and Kim et al. \cite{kim2018detecting}. Jung et al. \cite{jung2018malware} propose a reasonably smaller model than our models in terms of the number of parameters, but our model has a higher accuracy score than this model. Abijah et al. \cite{abijah2019intelligent}, Safa et al. \cite{safa2019benchmarking}, Kebede et al. \cite{kebede2017classification}, and Yan et al. \cite{yan2018detecting} propose deep learning models whose accuracy scores are close to our proposed models. In addition to these, Jang et al. \cite{jang2020fasttext} report five different accuracy scores for the BIG2015 dataset and they do not use F-Score despite the dataset is highly imbalanced. Their accuracy is the highest one for the dataset, but their network has two complex phases and uses GAN based data augmentation (they call this method malware obfuscator). On the other hand, RCNF does not use data augmentation, data resampling, transfer learning, and weighted loss functions for both datasets. In this perspective, RCNF is a simple version of an ensemble of CapsNet, and this simplicity highlights RCNF among its competitors. 


For those tables, the last studies using Malimg and BIG2015 datasets are chosen. To compare them fairly, these models are drawn from image-based malware analysis studies. These results show that while RCNF reaches the-state-of-the-art scores in terms of F-Score and accuracy on the Malimg dataset with less trainable parameter size, it outperforms its competitors on the BIG2015 with larger size of parameters. 

\section{Conclusion}
\label{sec:conc}
This paper introduces the first application of CapsNet on imbalanced malware family type classification task. Moreover, the first ensemble model of CapsNet called RCNF is introduced in this paper. The proposed models do not require any complex feature engineering methods or architecture for deep networks. To show that, we used two different malware family type datasets: Malimg and BIG2015. These datasets are used for image-based malware classification. Our proposed models can utilize these datasets directly using raw pixel values.

Datasets in the paper are highly imbalanced in terms of class distribution. CapsNet and RCNF do not use oversampling, under sampling, and weighted loss function during the training phase. Results show that CapsNet and RCNF are the best models least suffering from imbalanced class distribution among others in the literature.

Experiment results show that a single CapsNet model has good performance for both BIG2015 and Malimg datasets. However, we have assumed an ensemble model of CapsNet can help us to increase generalization performance and RCNF has better generalization performance results than a single CapsNet model as expected. In this point, results show that creating a bagging ensemble model CapsNet increases the performance on predicting rare malware classes. While single CapsNet can obtain $0.9779$ F-Score for the BIG2015 dataset, an ensemble of $10$ CapsNets achieves $0.9820$ F-Score. We can observe the similar effects on the Malimg dataset. It is shown that bagging increases the performance of the CapsNet for imbalanced datasets and the ensemble model is more successful at predicting rare classes than a single CapsNet model due to ensembling.  

Many models compared in this paper are designed complex and large in terms of the number of parameters. Some of them use data augmentation, weighted loss functions, different extra feature engineering methods, and pre-trained deep neural networks that have large number of parameters.

As for future work, we are planning to develop a hybrid architecture for malware classification. This hybrid method will be based on CapsNet architecture. 

\section*{Acknowledgement}
This work is supported by The Scientific and Technological Research Council of Turkey under the grant number 118E400.

\bibliographystyle{unsrt}
\bibliography{elsarticlenum.bib}







\end{document}